\documentclass[aps,pra,showpacs,twocolumn,superscriptaddress]{revtex4-1}
\usepackage{amsmath}
\usepackage{amsfonts}
\usepackage{graphicx}
\usepackage{dcolumn}
\usepackage{amsbsy}
\usepackage{bm}
\usepackage{amsthm}
\usepackage[caption=false]{subfig}
\usepackage{tabularx}
\usepackage{array}
\usepackage{natbib}

\begin{document}
\title{Optimizing classical communication in remote preparation of a general pure qubit}

\author{Congyi Hua}
 \affiliation{Zhejiang Institute of Modern Physics, Zhejiang University, Hangzhou 310027, China}
\author{Yi-Xin Chen}
 \email{yxchen@zimp.zju.edu.cn}
 \affiliation{Zhejiang Institute of Modern Physics, Zhejiang University, Hangzhou 310027, China}

\begin{abstract}
How to uses shared entanglement and forward classical communication to remotely prepare an arbitrary (mixed or pure) state has been fascinating quantum information scientists. A constructive scheme has been given by Berry for remotely preparing a general pure state with a pure entangled state and finite classical communication. Based on this scheme, for high-dimensional systems it is possible to use a coding of the target state to optimize the classical communication cost. Unfortunately, for low-dimensional systems such as a pure qubit the coding method is inapplicable. Because qubit plays a central role in quantum information theory, we propose an optimization procedure which can be used to minimize the classical communication cost in the remote preparation of a general pure qubit. Interestingly, our optimization procedure is linked to the uniform arrangement of $N$ points on the Bloch sphere, which provides a geometric description.
\end{abstract}
\maketitle

\section{Introduction}

In the field of quantum information processing, remote state preparation (RSP) is a kind of protocols that transmit a quantum state from a sender (``Alice") to a receiver (``Bob") using preshared entanglement and forward classical communication~\cite{Lo,Pati,Bennett}. Unlike the celebrated teleportation protocols~\cite{BennettTeleportation}, the sender does not possess a copy of the target state, but has complete classical knowledge of the state, which she chooses from a given ensemble. The RSP protocols can be divided into two different categories: exactly (non-asymptotically) faithful  and asymptotically faithful. The exactly faithful RSP produces the desired states one at a time, while the asymptotically faithful RSP only has an asymptotic efficiency. We are concerned with exactly faithful RSP in the present paper.

In the simple case where the target ensemble consists of a great circle on the Bloch sphere, the RSP can be done by using one maximally entangled state (ebit) and one classical bit (cbit) communication~\cite{Pati}. The constraint on the ensemble can be extended to the entire Bloch sphere by allowing more classical communication. Lo~\cite{Lo}, Leung and Shor~\cite{Leung} showed that two cbit communication is necessary and sufficient for the RSP of an arbitrary pure qubit with one ebit preshared. These investigations are based on Alice and Bob shared a maximally entangled state, however, the non-maximally entangled cases may occur due to the imperfect devices in the real world. In these cases, the required resource can be traded off between the cbits and ebits. Ye, \textit{et al.} proposed a protocol for remote preparation of an arbitrary pure state, by using finite cbits and non-maximally entangled pure state~\cite{Ye}. Soon a constructive scheme for this RSP protocol was given by Berry~\cite{Berry}.

Just as Berry showed, the classical communication cost increases drastically as the entanglement goes down. Although this is an inevitable consequence of the trade off between the two kind of resources, unnecessary classical communication cost should be minimized. For a large system dimension, Berry employed a coding of the target state to optimize the classical communication cost for the scheme of this type. However, for preparing low-dimensional target states such as a pure qubit, as the coding method is inapplicable, the scheme still suffers from unsatisfying classical communication cost. 

Since qubit is one of the central objects of study in quantum information theory, here we propose an optimization method which can be used in the remote preparation of a pure qubit. Our method comes from a rethink of the preliminary of Berry's scheme. We find the preliminary, which Berry described as an approximate scheme, is actually an algorithm for arranging points on the Bloch sphere. And changing the algorithm to one that can construct points distributed uniformly on the Bloch sphere will minimize the classical communication cost.

This paper is organized as follows. In Sec.~\ref{secii}, we restate Berry's scheme for preparing a general qubit as a four-step process.  We show how a uniform distribution of points on the Bloch sphere minimizes the classical communication cost for RSP scheme of this type. For clear demonstration of the optimization procedure, in Sec.~\ref{seciii} we introduce an algorithm called spiral points~\cite{Saff}, which can be used for easy construction of considerably uniformly distributed points on a sphere. Then we replace the original distributions in the scheme with the spiral points and compute the cbits versus ebits trade off. By comparing our results with those in Ref.~\cite{Berry}, we show that the cbits versus ebits trade off computed from the spiral points is very near a lower bound. Finally, in Sec.~\ref{seciv}, we summarize our results and draw some conclusions.

\section{Remote preparation of a general pure qubit}
\label{secii}
Berry's scheme aims at remote preparation of a pure state using any entangled pure state. Here we restate this scheme for preparing a general qubit.
 
Assume Alice and Bob share an entangled state, which has the form

\begin{equation}
\vert A\rangle =\sum _{k=0}^1 \alpha _k\vert k\rangle \vert k\rangle,
\label{eqsource}
\end{equation}
$ \alpha _k>0$, $\sum _{k=0}^1 \alpha _k^2=1$. Any two-qubit pure entangled state can be brought to this form via local unitary operations at Alice's location. The state Alice wants to prepare at Bob's side is denoted by $\vert \beta\rangle$, which is known to Alice but unknown to Bob.

Before we outline the procedure for Alice to remotely prepare $\vert \beta\rangle$, one important result from Ref.~\cite{Berry} need to be stated.  By Allowing Alice and Bob to perform local operations and communicate 2 bits of classical information, the possession of an entangled state in Eq.~(\ref{eqsource})  guarantees Alice the ability to remotely prepare an arbitrary qubit of the form 
\begin{equation}
\vert \psi\rangle =\sum _{k=0}^1 \psi _ke^{i \varphi _k}\vert k\rangle,
\label{eqcap}
\end{equation}
 where $\psi _0\geq 1-r^2$, $r=\min\{\alpha_i\}$.

 On the Bloch sphere, the ensemble of states that satisfy Eq.~(\ref{eqcap}) is represented by a $\vert 0\rangle$-centered spherical cap, denoted by $c_0$. According to the entanglement for pure qubits~\cite{Horodecki},  one can know that the less entanglement $\vert A\rangle$ has, the smaller spherical cap Alice can prepare.

Now let's outline the procedure for the preparation by four steps. To avoid unnecessary elaboration, we treat step 1 and 2 as briefly as possible. For more details, we refer the readers to Ref.~\cite{Berry}.

Step 0.  Construct a distribution of $N$ points (or states) $\vert\beta _i'\rangle$, $i=1, 2, \text{...}, N$, on the Bloch sphere. $N$ should be large enough to make the set of spherical caps $C=\left\{c_1, c_2, \text{...} , c_N\right\}$, where $c_i=\{\vert e\rangle\ \vert\ |\langle \beta _i'\vert e\rangle |^2\geq 1-r^2\}$, a cover of the Bloch sphere. Further, define $N$ unitary transformations $U_i$'s, each transforms $c_0$ into $c_i$.

Step 1-2. Alice prepares at Bob's location a state $\vert\varphi _0\rangle$ in $c_0$ such that Bob can bring $\vert\varphi _0\rangle$ to the desired state $\vert \beta\rangle$ in $c_i$ by some unitary transformation $U_i$. This can be done by an entanglement transformation followed by a disentangling measurement, and costs 2 bits of classical information.

Step 3. Alice send Bob $\log N$ bits classical information to indicate him which $U_i$ should be used to bring $\vert\varphi _0\rangle$ to $\vert \beta\rangle$.

Step 0 is actually the preliminary for the preparation in Berry's original scheme. We treat the preliminary as Step 0 to facilitate the explanation of the optimization procedure. In order to successfully perform the RSP with resource states having different entanglement, an algorithm should be given for constructing distributions of any number $N$ of points. It can be easily seen that the $\log N$ cbits cost in step 3 depends on the point distribution given in step 0.  If we have an algorithm can distribute the centers of the spherical caps more uniformly, then less number of spherical caps will be needed to cover the Bloch sphere, and more classical communication will be saved. To minimize the classical communication cost, what one need is an algorithm that constructs uniformly distributed points on the Bloch sphere. 

Below is how Berry's algorithm locates $N$ spherical caps. Assume the state locating at the center of a spherical cap is expressed as
\begin{equation*}
\vert\tilde{\beta '}\rangle =\sum _{k=0}^1 \beta _k\vert k\rangle,
\end{equation*}
where $\beta_0$ is real, and $\beta_1$ is complex. The state $\vert\tilde{\beta '}\rangle$ is not necessarily normalized and the corresponding normalized state will be denote by $\vert\beta '\rangle$. Berry begins with finding on the interval $\left[0,1\right]$ $D$ uniformly distributed numbers
\begin{equation*}
(2n-1)/D-1,
\end{equation*}
$n=1,2, \text{...} , D$. By picking 3 such numbers (repetition is allowed) as $\beta_0$, the real and imaginary parts of $\beta_1$, a spherical cap can be located. It's obvious that the total number of spherical caps constructed by this algorithm satisfies $N=D^3$.

\begin{figure}
	\includegraphics[width=0.36\textwidth]{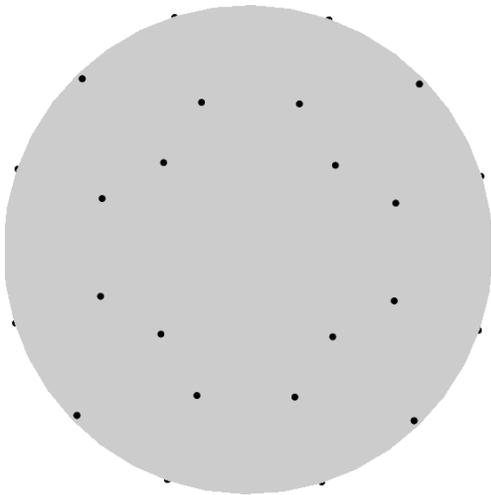}
	\caption{The points distribution given the Berry's algorithm in Ref.~\cite{Berry} in the case of $N=4^3$. Since some points coincide with each other, only 28 (instead of 64) points are distingushable. (View along the negative z direction.)}
	\label{fig1}
\end{figure}

In the above algorithm, although the spherical caps are represented by $N$ points that are uniformly distributed in the unit box, the distribution of these points on the Bloch sphere are nonuniform. Worse, as two or more different $\vert\tilde{\beta '}\rangle$'s may correspond to one the same $\vert\beta '\rangle$, lots of points coincide with each other. Fig.~\ref{fig1} illustrates the case of $N=4^3$.

We already know that the optimization procedure is equivalent to finding an algorithm for constructions of uniformly distributed points on the Bloch sphere. However except for some special cases such as the arrangements of 4, 8, 6, 12, 20 points on a sphere, in which cases we can use the vertices of the Platonic solids due to their perfect symmetry, finding an algorithm that can uniformly arrange an arbitrary number of points on a sphere is still an open question. Fortunately, there're still a variety of algorithms that can construct quite uniform point distribution on a sphere. A simple to describe and compute algorithm is spiral points, which we will use to demonstrate the optimization procedure in the Sec.~\ref{seciii}.

\section{Demonstrating the optimization procedure via spiral points}

\label{seciii}
The problem of how to uniformly distribute points on a sphere has long been receiving attention by scientists in their work, such as searching for large stable carbon molecules and locating identical charged particles so that they are in equilibrium according to Coulomb's law, etc. Spiral points is  an algorithm proposed for the explicit construction of considerably uniformly distributed points on the sphere. It has the advantage of being simple to describe and compute, thus suitable for the demonstration of the optimization procedure in remote preparation of a pure qubit.

\begin{figure}
	\includegraphics[width=0.36\textwidth]{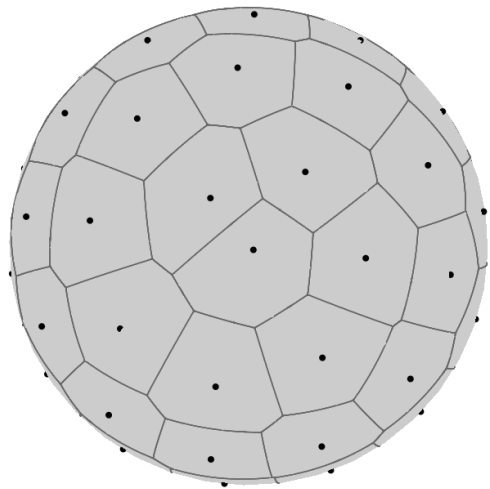}
	\caption{Spiral points for $N=64$. The mesh on the sphere shows the Voronoi cells corresponding to spiral points. (View along the negative z direction.)}
	\label{fig2}
\end{figure}

Just like the algorithm's name, the construction of the spiral points is like to draw a spiral path along the surface of the unit ball. One begins from setting the first spiral point at the south pole of the sphere. To obtain the next spiral point, one proceeds upward from the current point along a meridian to the height that is $2/(n-1)$ higher and travels counterclockwise along a latitude for a fixed distance of $3.6/\sqrt{N}$ to arrive at the next point. The entire path will end up at the north pole. Using spherical coordinates, the $i$th spiral points $p_i$ may be given as below: \begin{equation*}
\theta _i=\arccos \left(z_i\right),
\end{equation*}
\begin{equation*}
z_i=-1+\frac{2(i-1)}{N-1}
\text{, }
1\leq i\leq N
\text{, }
\end{equation*}
\begin{equation*}
\phi _1=\phi _N=0
\text{, }
\end{equation*}
\begin{equation*}
\phi _i=\left(\phi _{i-1}+\frac{3.6}{\sqrt{N}}\frac{1}{\sqrt{1-z_i^2}}\right)(\text{mod} 2\pi)\text{, }
2\leq i\leq N
\text{.}
\end{equation*}
In Fig.~\ref{fig2}, one can see how uniform the distribution of 64 spiral points looks.

Now let's calculate the cbits versus ebits trade off for the scheme using spiral points. But before we can calculate the trade of, we must introduce the concept of Voronoi diagram~\cite{Aurenhammer}. A Voronoi diagram is a way of dividing space into numbers of regions. In the context of Voronoi diagram the spiral points $p_i$'s are called sites. For each site, there will be a corresponding polygon-shaped region consisting of all points closer to this site than to any other. These regions are called Voronoi cells, whose edges are equidistant from two sites, and vertices equidistant from three or more sites. (Fig.~\ref{fig2} gives an illustration of Voronoi cells corresponding to the 64 spiral points.) Let's denote by $v_{\{i,j\}}$ the $j$th vertex of the Voronoi cell corresponding to $p_i$.

Since every spherical cap $c_i$ is centered at the spiral point $p_i$, $C$ will not become a cover of the Bloch sphere until every $c_i$ covers the Voronoi cell corresponding to $p_i$. One can measure the size of $c_i$ by the fidelity radius $r_F$, which is defined by 1 minus the fidelity between the central state and a boundary state, i.e.,  $r_F=r^2$. Similarly, one can measure the size of the hardest to cover Voronoi cell by
\begin{equation*}
\begin{split}&\rho _F(N)\\=&\max\left\{1-\left|\left\langle v_{\{i,j\}}\vert p_i\right\rangle \right|^2 \vert \text{ for } \text{any } p_i \text{ and } \text{related } v_{\{i,j\}}\text{}\right\}.
\end{split}
\end{equation*}
In order to make $C$ a cover of the Bloch sphere, $N$  should be large enough to ensure $\rho _F(N)\leq r_F$. To compute $\rho _F(N)$, we need to obtain the coordinates of all $v_{\{i,j\}}$'s. 

In the problems of generating a Voronoi diagram from a given set of points, except for some special points distributions, it is generally hard to find analytic solutions. One different approach that is commonly seen is to adopt a numerical solution. There're several algorithms developed for computing the spherical Voronoi diagram~\cite{Aurenhammer,Na,Zheng}. We implement in our program the popularly used sweep line algorithm for computing the Voronoi diagram in $O(N \log N)$ time~\cite{Zheng}. By taking the spiral points as input, the program is executed for the input size $N$ from 3 to 1024. We list part of the result (for $N=2^n$, $n=1, 2, ..., 10$) in the table below:

\begin{tabular*}{0.45\textwidth}{@{\extracolsep{\fill} } c|c|c|c|c|c }
  \hline
  $N$ & 2 & 4 & 8 & 16 & 32\\
  \hline
  $\rho_F$  & 0.5  & 0.5  & 0.259739 & 0.120679 & 0.054644\\
  \hline
 $N$ & 64 & 128 & 256 & 512 & 1024\\
  \hline
 $\rho_F$ & 0.026443 & 0.013054 & 0.006607 & 0.003326 & 0.001669\\
  \hline
\end{tabular*}

\begin{figure}
	\includegraphics[width=0.47\textwidth]{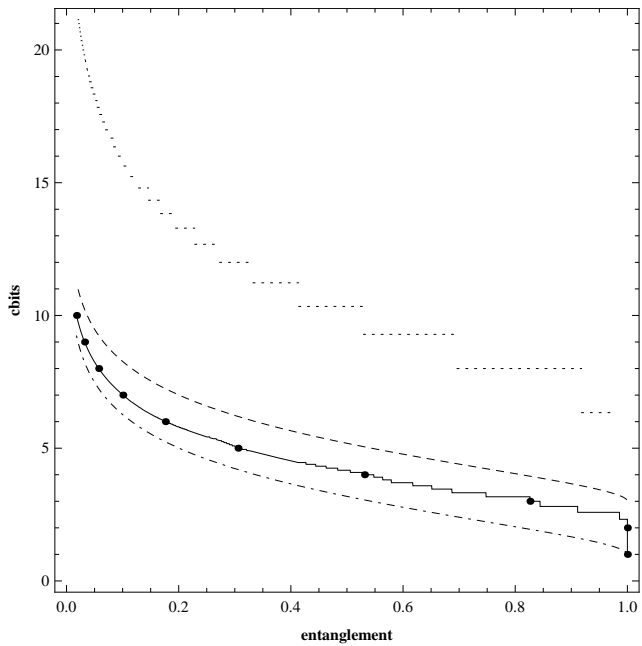}
	\caption{The cbits cost versus ebits for RSP of pure qubits states using partially entangled state. The dotted curve is that based on the original scheme given in Ref.~\cite{Berry}, and the solid curve is the result obtained when spiral points are used. The dashed-dotted curve and the dashed curve are an upper bound and a lower bound on the classical communication for RSP scheme of this type. The black dots are drawn from the cases which are presented in table.}
	\label{fig3}
\end{figure}

Based on the obtained values of $\rho _F(N)$, we can compute the classical bits cost versus entanglement of the resource state. The smallest integer $N$ which satisfies the inequation $r_F\geq \rho _F(N)$ is used to calculate the classical bits cost $\log N$ and the entanglement is calculated by $-r^2 \log r^2-(1-r^2)\log(1-r^2)$. We show the result in Fig.~\ref{fig3}. Comparing with that of the original scheme proposed in Ref.~\cite{Berry}, we can see that the classical bits cost after using spiral points is significantly reduced. Actually the classical bits cost is reduced to a level very close to the limit for RSP scheme of this type, because it is between an upper bound and a lower bound of this limit  (refer respectively to Eq. (23) and Eq. (24) of Ref.~\cite{Berry}).

It must be emphasized that Fig.~\ref{fig3} only plot the classical bits cost in step 3. The total classical bits cost for RSP schemes of this type should count the 2 bits in step 2 of the scheme.

\section{Conclusions}
\label{seciv}
We have reanalyzed a RSP scheme for remotely preparing a general pure state, and related the optimization of the scheme to an algorithm which can construct uniformly distributed points on the Bloch sphere. Since the original algorithm does not provide uniform point distributions on the Bloch sphere, we replace it with spiral points, an algorithm that gives a quite uniform point distribution with considerable simplicity.

Using a uniform point distribution algorithm like spiral points in the scheme has two main advantages.

(1) The classical bits cost of this type RSP scheme is reduced to a level near optimal, which the original scheme cannot achieve if the state to be prepared is in low dimension.

(2) Once an appropriate algorithm is determined, the scheme can be constructed easily. There is no need of a coding method to optimize the classical bits cost.

There may be some orther algorithms to choose, we use spiral points partly because its simplicity of describing and computing gives a good demonstration for the optimization procedure. To generalize our method to higher dimensions is possible although further work might be required.

\section{Acknowledgments}
This work is supported by the NNSF of China, Grant No. 11375150.

\end{document}